\documentclass{article}
\usepackage[T1]{fontenc}
\usepackage{wrapfig}
\usepackage[portuges,english]{babel}
\usepackage{amssymb,latexsym,amsmath,color,mathrsfs,ifsym,graphics,stmaryrd} 
\usepackage[colorlinks,linkcolor=blue,urlcolor=blue,citecolor=black,
plainpages=false,pdfpagelabels,breaklinks]{hyperref}

\title{QBism, FAPP and the Quantum Omelette.\\(Or, Unscrambling Ontological Problems\\ from Epistemological Solutions in QM)}
\author{{\sc Christian de Ronde}\thanks{Fellow Researcher of the Consejo
Nacional de Investigaciones Cient\'{\i}ficas y T\'ecnicas.}}
\date{\begin{center}
\begin{small}
Philosophy Institute Dr. A. Korn \\ 
Buenos Aires University, CONICET - Argentina \\
Center Leo Apostel and Foundations of  the Exact Sciences\\
Brussels Free University - Belgium \\
\end{small}
\end{center}}

\usepackage[margin=3.5cm]{geometry}

\begin{document}
\maketitle

\bigskip

\bigskip

\begin{abstract}
\noindent In this paper we discuss the so called ``quantum omelette'' created by Bohr and Heisenberg through the mix of (ontic) objective accounts and (epistemic) subjective ones within the analysis of Quantum Mechanics (QM). We will begin by addressing the difficult relation between ontology and epistemology within the history of both physics and philosophy. We will then argue that the present ``quantum omelette'' is being presently cooked in two opposite directions: the first scrambling ontological problems with epistemological solutions and the second scrambling epistemic approaches with ontological questions. A good example of the former is a new type of argumentation strategy attempting to justify the use of decoherence, namely, the ``For All Practical Purposes'' (shortly known as FAPP) type of justification. We will argue that `FAPP-type solutions' remain, at best, epistemological answers which not only escape the ontological questions at stake ---regarding the quantum to classical limit--- but also turn the original problem completely meaningless. The latter omelette can be witnessed in relation to some criticisms raised against the epistemic Bayesian approach to QM (shortly known as QBism). We will argue that QBists have produced a consistent scheme that might allow us to begin to unscramble ---at least part of--- the ``quantum omelette''. In this respect, we will show why the epistemic QBist approach is safe from several (ontological) criticisms it has recently received (see \cite{Marchildon15, Mohrhoff14, Nauenberg15}). We end our paper with a discussion about the importance of ontological approaches within foundations of QM.
\medskip
\end{abstract}

\bigskip

\bigskip

\textbf{Keywords}: QBism, FAPP, Ontology, Epistemology, Quantum Mechanics.

\renewenvironment{enumerate}{\begin{list}{}{\rm \labelwidth 0mm
\leftmargin 0mm}} {\end{list}}

\newcommand{\ita}{\textit}
\newcommand{\mcal}{\mathcal}
\newcommand{\mfrak}{\mathfrak}
\newcommand{\mbb}{\mathbb}
\newcommand{\mrm}{\mathrm}
\newcommand{\msf}{\mathsf}
\newcommand{\mscr}{\mathscr}
\newcommand{\lra}{\leftrightarrow}
\renewenvironment{enumerate}{\begin{list}{}{\rm \labelwidth 0mm
\leftmargin 5mm}} {\end{list}}

\newtheorem{dfn}{\sc{Definition}}[section]
\newtheorem{thm}{\sc{Theorem}}[section]
\newtheorem{lem}{\sc{Lemma}}[section]
\newtheorem{cor}[thm]{\sc{Corollary}}
\newcommand{\Proof}{\textit{Proof:} \,}
\newcommand{\cqd}{{\rule{.70ex}{2ex}} \medskip}

\bigskip

\newpage

\section*{Introduction}

This paper focusses on what has been called by Jaynes the ``quantum omelette'' \cite{Jaynes}, an improper scrambling of (ontic) objective and (epistemic) subjective perspectives in the analysis of Quantum Mechanics (QM) that has been taking place since the early debates of the founding fathers. In order to discuss the problematic relation between ontology and epistemology in QM we begin, in section 1, by a rough review of some important elements, present within the history of physics and philosophy, which regard the fundament, meaning and reference of physical theories. This introduction attempts to distinguish two main lines of thought which have been in tension since the origin of Western thought. These two perspectives, as we shall discuss through the paper, are still present in the contemporary debates about the meaning and reference of quantum theory itself. In section 2 we discuss the many problems of the theory of quanta in order to provide an objective representation of physical reality. Section 3 relates the origin of the so called ``quantum omelette'' to the early discussions of Einstein, Bohr and Heisenberg. In section 4, we will argue that a good example of the contemporary scrambling of the omelette is a recent type of solution introduced in the foundational literature attempting to justify the principle of decoherence called: ``For All Practical Purposes'' (shortly known as FAPP). We will argue that `FAPP-type solutions' remain, at best, epistemological answers which not only escape the ontological questions at stake (regarding the quantum to classical limit) but also turn the original problem completely meaningless. Section 5 introduces the Bayesian approach to QM (shortly known as QBism) as a consistent epistemic scheme that might allow us to begin to unscramble ---at least part of--- the ``quantum omelette''. In this respect, in section 6, we will show why QBism is completely safe from several (ontological) criticisms it has recently received (see \cite{Marchildon15, Mohrhoff14, Nauenberg15}). In section 7 we discuss how the QBist approach dissolves ontological problems. Section 8 argues in favor of the importance of ontological problems within foundations of QM. Finally, recalling our philosophical introduction, we end section 9 with some final philosophical remarks.

\section{Philosophy, Physics and Sophistry}

It is widely accepted that the origin of Western thought goes back to Ancient Greece. Around the 7th Century B.C., between the beautiful islands of the Aegean sea, some of their inhabitants started to claim and argue in favor of a powerful idea. That the world was not commanded by the desires and wishes of the Gods, but by certain specific rules and laws that the whole Cosmos was obliged to follow. Furthermore, they claimed that it was possible to learn the true fundament of such Cosmos, called {\it physis}.\footnote{The term {\it physis} is a Greek theological, philosophical, and scientific term usually translated into English as ``Nature''.} They argued it was possible to create {\it theories} that could explain what happened in Nature, how things {\it changed} in the world, why the stars moved in the sky or how birds could fly. This group of thinkers, who claimed to be in love {\it episteme} (true knowledge), who and argued that the fundament existence and reality was {\it physis}, called themselves ``philosophers'' or ``physicists''. 

Philosophers, who claimed to love knowledge in itself, started the first battle of Western thought when they accused the sophists to be merely interested in selling their rhetoric knowledge in the streets of the Greek {\it polys}, corrupting the young with their teachings about how to turn mere {\it doxa} into a triumph with no fundament. Indeed, according to the philosophers, the sophists were not searching for an understanding of the world, they were only interested in winning discussions through rhetoric argumentation. Sophists trained their students in order to triumph in the Agora beyond being right or wrong. Protagoras dictum that ``man is the measure of all things'' made explicit the fact that sophists were not interested in the problems raised by the new tribe of thought. While sophists accepted the {\it simulacra}, the many faces and masks of existence, philosophers begun one of the most outstanding quests in human history, a search for true knowledge about the world and reality. 

It was Plato, with his dialogues who begun a history of victory for philosophy against sophistry. Since then, philosophy ---in deep relation to metaphysics--- has witnessed a wild search for the true fundament of reality and existence. However, independently of the many metaphysical schemes developed ever since Plato, we could remind the words of Alfred North Whitehaed \cite{Whitehead} who argued: ``[t]he safest general characterization of the European philosophical tradition is that it consists of a series of footnotes to Plato.'' Indeed, Plato's basic metaphysical scheme was repeated once and again, changing only that which served as the fundament within each new metaphysical system. Instead of a world of ideas, Aristotle presented his hylomorphic scheme in which the potential and actual realms related through {\it dynamis}. While Spinoza developed his notion of {\it substance}, Leibniz presented an architectonic founded on {\it monads}, Schopenahuer created the notion of {\it will} that would confront the {\it representation of the world}, and Hegel developed in the same period his dialectics and the notion of {\it absolute concept}. Of course the list continues, but what is common to them all, is the idea that existence and reality can be philosophically represented through a metaphysical system. While metaphysical schemes create their own fundament, all of them serve the same purpose: {\it to represent reality}.   

Protagoras dictum found a new version as late as the 17th Century, when Ren\'e Descartes attempted to fundament knowledge in terms of thought and reason. {\it Cogito ergo sum} [I think, thus I exist]. The subject was a creation of modernity, the new fundament of a time more finite, maybe less ambitious than that of the Greeks. After David Hume's critic to the inductive nature of science and the impossibility to ground the notion of causation in experience, the knowledge provided by physical theories had been relegated to a mere {\it habit}. It was Immanuel Kant ---a physicist himself--- who through the creation of a {\it transcendental subject} managed to justify Newtonian physics in terms of (objective) knowledge. The price to pay for the new Kantian physics was the abandonment of its direct relation to {\it Das Ding an sich} [Reality {\it as it is}]. Hence, the reality of the world was replaced by the less ambitious {\it objective reality} of subjects. Physics had to abandon the Greek {\it episteme} and be content with a new type of certainty: {\it objective knowledge}. A finite knowledge with which we humans shaped experience.

With Kant, for the first time, it was understood that our (categorical) representation of reality ---given by the (mainly Aristotelian) categories and the forms of intuition (Newtonian space and time); both analyzed in the {\it Critique of Pure Reason}--- limits and configures in a definite manner the objects around us. Causality, identity, non-contradiction, etc., were not Platonic concepts or ideas that humans had {\it discovered} in a heaven populated by them, but rather the {\it a priori} conditions of human understanding itself. This distance between our categorical representations and phenomena opened the door for thinking about such categories in terms of {\it creations} ---maybe too human. The problem of representation was born: how can we relate our (internal) representation with the real (external) world?\footnote{This problem was, of course, present within two main philosophical positions before Kant. Rationalism, with Descartes as its major figure, ended up securing the relation between cogito and reality through the good consciousness of God. The empiricist has exactly the inverse problem in order to describe sensible empiria without reference to metaphysical notions such as identity, causality, etc. In the context of the positivist tradition this problem was revived in the sixties and seventies in terms of the ``theory ladenness of physical observation''.} The world and reality had begun very slowly to dissolve. 

In the second half of the 19th Century, by imposing a reconsideration of observability beyond the {\it a priori} categories of Kantian metaphysics, Machian positivism begun to deconstruct the fundaments of classical Newtonian physics. It was this subversive analysis that forged the key to unlock a new physics, a new experience. Classical Newtonian concepts had become dogmatic {\it a priori} notions within Kantian metaphysics. The positivist deconstruction of Newtonian mechanics ---on which the Kantian architectonic was grounded--- produced one of the most important crisis in human thought. However, it was this same crisis that provided the conditions of possibility for the development of new physical theories. These theories went far beyond the limits imposed by classical notions. At the beginning of the 20th Century the crisis imposed by the positivistic critic to classical physics became the soil on which two of the most amazing theories ever imagined started slowly to grow. We are talking here of course about Relativity Theory and QM. 

For many centuries, since the battles of the Greeks, the philosophers and physicists ruled the world of metaphysical thought. But the sophists were not dead, they were just silently awaiting for better times to come. They had to wait long, but the time would come. It was as late as the 20th Century that the sophist flag was raised once again. Logical positivists, following Mach, had fought strongly against dogmatic (Kantian) metaphysical thought. Against {\it a priori} metaphysical concepts, they argued in their famous {\it Manifesto} \cite{VC}: ``Everything is accessible to man; and man is the measure of all  things. Here is an affinity with the Sophists, not with the Platonists; with the Epicureans, not with the Pythagoreans; with all those who stand for earthly being and the here and now.'' Their main attack to metaphysics was designed through the idea that one should focus in ``statements as they are made by empirical science; their meaning can be determined by logical analysis or, more precisely, through reduction to the simplest statements about the empirically given.'' The positivist architectonic stood on the distinction between {\it empirical terms}, the empirically ``given'' in physical theories, and {\it theoretical terms}, their translation into simple statements. Such separation and correspondence between theoretical statements and empirical observation would have deep consequences, after the second world war, not only regarding the problems addressed in philosophy of science but also with respect to the limits of development of many different lines of research within QM itself.\footnote{As remarked by Curd and Cover: ``Logical positivism is dead and logical empiricism is no longer an avowed school of philosophical thought. But despite our historical and philosophical distance from logical positivism and empiricism, their influence can be felt. An important part of their legacy is observational-theoretical distinction itself, which continues to play a central role in debates about scientific realism.'' \cite[p. 1228]{PS}}

In conclusion, since the 5th Century B.C., both physics and philosophy had ruled the world of thought. All physical theories had been understood as describing or representing physical reality, as relating to Nature, as referring to the Cosmos and existence, to {\it physis}. This was until the 20th Century, when QM changed everything...

\section{Quantum Mechanics: Physics and Reality}

Physics originated itself from what might be regarded as a naive idea of clever animals: the world and reality {\it exist}. Reality {\it is}. The world and reality stand before us as an amazing playground. As Albert Einstein \cite{Einstein79} expressed it most beautifully: ``Out yonder there was this huge world, which exists independently of us human beings and which stands before us like a great, eternal riddle, at least partially accessible to our inspection.'' Indeed, after Darwin it could be claimed ---escaping religious thought--- that humans are only animals, very clever and powerful animals. Humans are so powerful animals, that are even capable of destroying the whole world. But still we are only that, not more, not less: existents within Nature. Existents that ---after Hume and Kant--- must recognize, shape experience not only through their senses but also through their metaphysical and categorical presuppositions. Causation, as Hume clearly exposed, is not something empirically grounded, it is never found in the observable world. Rather, as Kant would later on remark, it is a metaphysical presupposition which allows the subject to make sense of observations. In particular, identity and non-contradiction are not principles that we find in Nature. Quite the opposite, they are the very conditions that define and constrain our experience of Nature.

The multiple representations provided by different theories in the 20th Century might suggest that as much as {\it discoveries} they might have seemed to be, the conditions of human understanding are also {\it creations}.\footnote{Of course a hard core Pythagorean-Platonist could argue against this position and claim that mathematics is in fact related to reality. There is a whole literature which presupposes and debates about such a relationship. Contrary to this metaphysical position, we consider mathematics as a non-representational discipline which is in no way constrained by metaphysical principles or the real world ---whatever that might be. It is only physics, which making use of mathematical formalisms, attempts to discuss about physical reality.} As remarked by Wolfgang Pauli \cite[p. 95]{Pauli94} when discussing the breakdown of the fundaments in the 20th Century: ``The modern physicist regards with skepticism philosophical systems which, while imagining that they have definitively recognized the {\it a priori} conditions of human understanding itself, have in fact succeeded only in setting up the {\it a priori} conditions of the systems of mathematics and the exact sciences of a particular epoch.'' In this respect, Whitehead would make the strong point that, since our theories and representations are human creations, we are always ---almost exclusively--- confronted with ourselves:

\begin{quotation}
\noindent {\small``We have found a strange footprint on the shores of the unknown. We have devised profound theories, one after another, to account for its origins. At last, we have succeeded in reconstructing the creature that made the footprint. And lo! It is our own.'' \cite[pp. 200-201]{Whitehead}} \end{quotation} 

\noindent But, are physical theories only that?  Solipsistic creations? Are physical theories only mirrors that unveil our own reflection? These questions places us at a crossroad. 

Within physical discourse one of the basic cornerstones is simple down to earth counterfactual reasoning. I am talking here of the intuitive idea which any physicist assumes as a basic presupposition of the discipline. According to this idea, given an empirically adequate physical theory I can produce (counterfactual) statements regarding phenomena which no physicist will ever doubt. We physicists know that a small ball will fall accelerated at 9.8 $\frac{m}{s^2}$ here on Earth. And we also know that the same ball will fall accelerated at 1.6 $\frac{m}{s^2}$ in the moon. We know how it would fall in Jupiter or any other distant planet, even though we might never reach them in order to actually perform the experiment. We know how it would fall in an imaginary planet of a specific mass and radius. This is the reason why, counterfactual reasoning and discourse are the cornerstones of conceptual physical representation itself. Indeed, the intuition of physicists is also related to the possibility of prediction. As expressed by Robert Griffiths:

\begin{quotation}
\noindent {\small ``If a theory makes a certain amount of sense and gives predictions which agree reasonably well with experimental or observational results, scientists are inclined to believe that its logical and mathematical structure reflects the structure of the real world in some way, even if philosophers will remain permanently skeptical.'' \cite[p. 361]{Griffiths02}} \end{quotation}

The classical representation of physics was produced after many centuries, when Newton related the theory of calculus with a meta-physical view of the world in terms of physical notions such as: `space', `time', `particle', `force', `mass', etc. This view was extended by Maxwell electromagnetic equations and the creation of new physical notions such as that of `field' and `charge'. These physical theories allowed us to {\it represent} what the world is like. 

QM was born from the dissolution of our classical representation of the world. Planck's quantum postulate ---which introduced the {\it discrete} nature of quanta--- lies at the origin of such defoundation. Few decades after Planck's postulate, Heisenberg advanced the first closed formalization of the theory by escaping Bohr's ``magical'' model of the atom ---as Sommerfeld used to call it. Returning to the positivist methodological (Machian) rule according to which, only `observable magnitudes' should be considered within a theory, he was able to develop matrix mechanics. The quantum theory was not designed to talk about ``trajectories of particles'', it was constructed from the very departure of such (classical) (meta-)physical representation. Einstein's dictum: ``it is only the theory which can tell you what can be observed'', guided Heisenberg in order to further advance in the derivation of the {\it indeterminacy principle} from the exclusive use of the quantum postulate and matrix mechanics. The new quantum formalism had many surprises to give to those physicists which expected to understand QM in terms of ``classical reality''. Very soon, it also became clear, as remarked by Dirac \cite[p. 12]{Dirac74}, that ``[t]he nature of the relationships which the superposition principle requires to exist between the states of any system is of a kind that cannot be explained in terms of familiar physical concepts.'' Schr\"odinger himself made explicit, through an {\it ad absurdum} proof, the inadequacy of classical concepts to account for these strange mathematical elements of the theory \cite{Schr35}.   

The departure of QM from our classical world view had two main reactions. The first was an  attempt to ``complete the theory'' with (hidden) variables that would allow us to restore a classical understanding about {\it what there is}. The second reaction, endorsed ---for not so different reasons--- by Bohr and the logical positivists, was the abandonment of the physical representation of QM itself. Both reactions became the main lines of research within the foundational investigations about QM in the second half of the 20th Century, and are still today firmly in place.

\section{Bohr, Heisenberg and the Quantum Omelette}

An ontological question is a question about the nature of being, existence and reality. An ontological question presupposes reality and the possibility to represent it in some way. From an ontological perspective, epistemological questions related to the acquisition of knowledge by subjects are of secondary importance. There is a deep and obvious difference between, discussing about `what reality is', and discussing about `how subjects are able to acquire knowledge from experience'. From an ontological perspective, humans are completely superfluous, we could be simply out of the picture. The reason is that we assume a perspective from which we are not so different from any other existent within Nature. We are just {\it as} important as anything else. 

An epistemological answer ---as understood today--- relates to the way in which we humans (also called `subjects', `agents', `users', `persons', etc.) acquire knowledge. Hence, epistemology assumes a different perspective from ontology, focusing in how subjects relate to knowledge and experience. Human beings and their observations are in this case the point of departure. However, it can be also argued that epistemic philosophical perspectives are just another footnote to Plato, another scheme in which the fundament is ---instead of the Greek {\it physis}--- the subject itself. 

The introduction of the subject in physics begun in modernity with Kant's critic to dogmatic metaphysics and the acceptance of our finitude. In order to justify the objective character of Newtonian mechanics, humans had to enter the scene of physical representation in an intrinsic manner. Kant's architectonic found an answer to Hume's critic to physics by placing the subject as the cornerstone itself of physical understanding. Since then, Western thought started to abandon the search for truth and (infinite) reality, a quest regarded ---not without reason--- as a suspicious enterprise. From a more sober epistemological perspective, (external) reality couldn't  be regarded anymore as the ground or goal of understanding itself. The notion of `reality' which had been always understood as independent of subjects had begun ---within Kantian metaphysics--- to get mixed with them. 

Epistemic approaches do not seek necessarily for a {\it referent} beyond observation itself, and in this sense their accounts need not justify their empirical findings. Indeed, epistemic approaches remain safe when they restrict their discourse to the way we humans (subjects, persons, users, etc.) interact through the mutual communication of such empirical findings; when they leave on the side the relation of these interactions to the world and reality themselves; when they remain on the surface of {\it intersubjectivity}.\footnote{As we shall see, the notion of intersubjectivity is key to Bohr's approach to QM.} By denying the need of providing a reference to their empirical grounding epistemological perspectives escape the problem of reality. Physical theories are then transformed into mere algorithmic devices, ``economies of (human) experience'' ---as Mach used to characterize physical theories. In this way all ontological questions about the fundament of such relations are dissolved, or ---in the worst of cases--- considered simply as ``metaphysical bla bla''. 

The choice between a philosophy that attempts to address the question of (infinite) reality and existence, and one that prefers to ground itself on the supposedly less ambitious task of understanding how (finite) humans relate to experience, is just that, a philosophical choice of how to approach certain specific problems and questions. 

The philosophical stance that we assume defines the specific problems, the possible questions and (even) answers that fall within our system of thought. But a limit is also a possibility, an horizon. Problems are not ``out there'', they are part of a definite viewpoint with definite metaphysical assumptions, presuppositions, without which they cannot be even stated. This is why, we are close to van Frassen \cite[p. xviii]{VF08} when he shouts: ``I argue for a view of philosophy as a stance, as existential.'' Indeed, as he explains: 

\begin{quotation}
\noindent {\small``Philosophy itself is a value- and attitude-driven
enterprise; philosophy is in false consciousness when it sees itself
otherwise. To me philosophy is of overriding importance, to our
culture, to our civilization, to us individually. For it is the
enterprise in which we, in every century, interpret ourselves anew.
But unless it so understands itself, it degenerates into an arid
play of mere forms.'' \cite[p. 17]{VF08}} \end{quotation}
 
Even though for many philosophers this introduction might sound as a set of obvious remarks, they are of outmost importance in order to understand the most weird state of affairs occurring today when addressing philosophical and foundational issues about QM. Indeed, within the philosophy of QM we are at a stage where ontology and epistemology have been mixed up in an omelette that we need to unscramble. As Jaynes makes the point:

\begin{quotation}
\noindent {\small``[O]ur present [quantum mechanical] formalism is not purely epistemological; it is a peculiar mixture describing in part realities of Nature, in part incomplete human information about Nature ---all scrambled up by Heisenberg and Bohr into an omelette that nobody has seen how to unscramble. Yet we think that the unscrambling is a prerequisite for any further advance in basic physical theory. For, if we cannot separate the subjective and objective aspects of the formalism, we cannot know what we are talking about; it is just that simple.''  \cite[p. 381]{Jaynes}}
\end{quotation}

\noindent Simply put: if we assume an ontological perspective there can be no {\it choice} which determines what reality {\it is}. We humans and our choices have to be out of the picture, not because we are good or bad, but because otherwise physical representation becomes impossible. A subject can not define, within a particular physical representation of a theory, what is physically real through `a choice'. Physical reality can be only represented in an objective manner if the subject plays no essential role within that representation. A subject-choice-dependent reality is not an interesting reality for a physicist. If choice enters the scene in the determination of physical reality then there is no possibility of having a physical representation of reality which goes beyond the {\it hic et nunc}. And exactly that, going beyond the here and now, is the main power of physics.

Einstein showed the inconsistencies with respect to physical reality in which QM had been drawn through Bohr's complementarity approach. In the EPR paper \cite{EPR}, he defined his now famous {\it elements of physical reality}. Regardless of the specificity of Einstein's definition, in his reply Bohr simply evaded the question at stake \cite{Bohr35}. Bohr did not answer the main problem: what should be considered physically real according to QM? Instead, he shifted the debate focusing on the problem of complementary measurements making a long exposition of classical experimental arrangements [{\it Op. cit.}, pp. 697-9]. The ``essential ambiguity'' ---as Bohr called it--- in Einstein's definition of what had to be considered as physically real was not further developed nor addressed. It was not even replaced by a different, more adequate notion, it was simply left aside. Instead, Bohr explained how things {\it had to be done}; how, following a set of rules, one could recover from QM a ``rational account of classical physical phenomena'' \cite{BokulichBokulich}. This ``rational account'' was clearly not an ontological account, it was an epistemic one. Bohr understood very well he could not talk about physical reality within his complementarity scheme, for that would doom his system incoherent. An object cannot be represented by two mutually incompatible representations such as that of `wave' and `particle'. That is the reason why Bohr always talked about ``knowledge'' ---evading physical reality--- and added in every phrase an ``as if'' providing a ``foggy reference'' to such classical representations \cite{deRonde15}. 

Bohr's epistemological approach carefully escapes ontological debates. His scheme was consistent and difficult to tackle. Even today it seems to us one of the strongest approaches to QM. According to his long time assistant, Aage Petersen, when asked whether the quantum theory could be considered as somehow mirroring an underlying quantum reality, Bohr \cite[p. 8]{WZ} declared: ``There is no quantum world. There is only an abstract quantum physical description. It is wrong to think that the task of physics is to find out how nature is. Physics concerns what we can say about nature.'' Petersen himself makes clear the distance of Bohr with respect to ontological concerns. 

\begin{quotation}
\noindent {\small ``Traditional philosophy has accustomed us to regard
language as something secondary and reality as something primary.
Bohr considered this attitude toward the relation between language
and reality inappropriate. When one said to him that it cannot be
language which is fundamental, but that it must be reality which, so
to speak, lies beneath language, and of which language is a picture,
he would reply, ``We are suspended in language in such a way that we
cannot say what is up and what is down. The word `reality' is also a
word, a word which we must learn to use correctly'' Bohr was not
puzzled by ontological problems or by questions as to how concepts
are related to reality. Such questions seemed sterile to him. He saw
the problem of knowledge in a different light.'' \cite[p. 11]{Petersen63}}
\end{quotation}

\noindent Indeed, Bohr's own characterization of physics goes in line with such departure from ontology and his emphasis on human experience and communication:

\begin{quotation}
\noindent {\small ``Physics is to be regarded not so much as the study of
something a priori given, but rather as {\small{\it the development
of methods of ordering and surveying human experience.}} In this
respect our task must be to account for such experience in a manner
independent of individual subjective judgement and therefor
{\small{\it objective in the sense that it can be unambiguously
communicated in ordinary human language.}}'' 
\cite{Bohr60} (emphasis added)} \end{quotation}

\noindent As it becomes clear, the price Bohr was willing to pay was the abandonment of the direct relation between physics and reality. 

\begin{quotation}
\noindent {\small ``I am quite prepared to talk of the spiritual life of an
electronic computer; to say that it is considering or that it is in
a bad mood. What really matters is the unambiguous description of
its behavior, which is what we observe. The question as to whether
the machine {\small {\it really}} feels, or whether it merely looks
as though it did, is absolutely as meaningless as to whether light
is `in reality' waves or particles. {\small{\it We must never
forget that `reality' too is a human word just like `wave' or
`consciousness.' Our task is to learn to use these words correctly
---that is, unambiguously and consistently.}}'' \cite[p. 5]{WZ} (emphasis added)} \end{quotation}

\noindent This was the line of development of a great part of physics in the second half of the 20th Century, including QM. Indeed, as remarked by Arthur Fine: 

\begin{quotation}
\noindent {\small ``[The] instrumentalist moves, away from a realist construal of the emerging quantum theory, were given particular force by Bohr's so-called `philosophy of complementarity'; and this nonrealist position was consolidated at the time of the famous Solvay conference, in October of 1927, and is firmly in place today. Such quantum nonrealism is part of what every graduate physicist learns and practices. It is the conceptual backdrop to all the brilliant success in atomic, nuclear, and particle physics over the past fifty years. Physicists have learned to think about their theory in a highly nonrealist way, and doing just that has brought about the most marvelous predictive success in the history of science.'' \cite[p. 1195]{PS}}
\end{quotation}

The triumph of the Danish physicist can be witnessed in the fact that the so called ``Copenhagen interpretation'' ---which has many of Bohr's main ideas at play--- is taught in all Universities around the globe. But even though Bohr presented an epistemic approach which attempted to escape an ontological reference of the quantum formalism, there were nevertheless, within Bohr's interpretation, strong metaphysical presuppositions at play. In fact, there are three important Bohrian (metaphysical) ideas which have turned into dogma in the present literature regarding foundational issues about QM. The first (metaphysical) presupposition is the idea that there must exist a ``quantum to classical limit'' ---assuming what Bokulich calls an ``open theory approach'' \cite{Bokulich04}---, the principle that one must find a ``bridge'' or ``limit'' between classical mechanics and QM.\footnote{This idea can be related to Bohr's correspondence principle and his attempt to provide a rational account of QM in terms of classical physics. See for discussion: \cite{BokulichCP, BokulichBokulich}.} The second (metaphysical) presupposition is the idea that classical physical language is a {\it necessary condition} for discussing about all physical phenomena and experience. 

\begin{quotation}
\noindent {\small ``It is decisive to recognize that, however far the phenomena transcend the scope of classical physical explanation, the account of all evidence must be expressed in classical terms. The argument is simply that by the word `experiment' we refer to a situation where we can tell others what we have done and what we have learned and that, therefore, the account of the experimental arrangement and of the results of the observations must be expressed in unambiguous language with suitable application of the terminology of classical physics.'' \cite[p. 209]{Bohr49}}
\end{quotation}

\noindent According to Bohr \cite[p. 7]{WZ}: ``[...] the unambiguous interpretation  of any measurement must be essentially framed in terms of classical physical theories, and we may say that in this sense the language of Newton and Maxwell will remain the language of physicists for all time.'' Closing the possibility of creating new physical concepts, Bohr [{\it Op. cit.}] argued that ``it would be a misconception to believe that the difficulties of the atomic theory may be evaded by eventually replacing the concepts of classical physics by new conceptual forms.'' The third presupposition relates the previous ones through an empiricist condition: the analysis of QM must begin from the analysis of (classical) measurement observations and not from the mathematical formalism of the theory. 

These three Bohrian dictums, properly mixed with a realist flavor have created a set of ``no-problems'' about QM which have been intensively discussed for many decades:  non-separability, non-individuality, non-locality, non-distributivity, non-identity, etc. These problems impose implicitly ``right from the start'' a set of classical (metaphysical) presuppositions within QM, namely, separability, individuality, locality, distributivity, identity, etc. All of them are ontological (metaphysical) problems which attempt to discuss and analyze the reference of the quantum formalism with respect to (classical) physical reality. But there are, in between them, two most interesting problems in which the intrusion of a choosing subject appears explicitly in the determination of what is considered to be (classically) real ---or actual. 

The first of these problems is the so called ``basis problem'' which attempts to explain how is Nature capable of making a choice between different incompatible bases. Which is the objective physical process that leads to a particular basis instead of a another one?  If one could explain this path through an objective physical process, then the choice of the experimenter could be regarded as well as part of an objective process ---and not one that determines reality. Unfortunately, still today the problem remains with no solution within the limits of the orthodox formalism. There is no physical representation of the process without the addition of strange {\it ad hoc} rules (see \cite{deRonde16}). 

The second is known infamous ``measurement problem''.  Given a specific basis (context or framework), QM describes mathematically a state in terms of a superposition (of states). Since the evolution described by QM allows us to predict that the quantum system will get entangled with the apparatus and thus its pointer positions will also become a superposition,\footnote{Given a quantum system represented by a superposition of more than one term, $\sum c_i | \alpha_i \rangle$, when in contact with an apparatus ready to measure, $|R_0 \rangle$, QM predicts that system and apparatus will become ``entangled'' in such a way that the final `system + apparatus' will be described by  $\sum c_i | \alpha_i \rangle  |R_i \rangle$. Thus, as a consequence of the quantum evolution, the pointers have also become ---like the original quantum system--- a superposition of pointers $\sum c_i |R_i \rangle$. This is why the {\it MP} can be stated as a problem only in the case the original quantum state is described by a superposition of more than one term.} the question is why do we observe a single outcome instead of a superposition of them? 

It is interesting to notice that for Bohr, the measurement problem was never considered. The reason is that through his presuppositions, Bohr begun the analysis of QM presupposing ``right from the start'' classical single outcomes. Bohr evaded in this way the discussion about the meaning and representation of quantum superpositions (see for a detailed analysis \cite{deRonde16c}). As explained by Dieks: 

\begin{quotation}
\noindent {\small ``We thus have returned to the theme of the indispensability of classical concepts [in the Bohrian scheme]: measuring devices, like all macroscopic objects around us, can and must be described classically. It is an immediate consequence of this that measurements necessarily have only one single outcome. Pointers can only have one position at a time, a light flashes or does not flash, and so on ---this is all inherent in the uniqueness of the classical description. Because of this, Bohr?s interpretation does not face the ``measurement problem'' in the form in which it is often posed in the foundational literature, namely as the problem of how to explain ---in the face of the presence of superpositions in the mathematical formalism--- that there is only one outcome realized each time we run an experiment. For Bohr this is not something to be explained, but rather something that is given and has to be assumed to start with. It is a primitive datum, in the same sense that the applicability of classical language to our everyday world is a brute fact to which the interpretation of quantum mechanics necessarily has to conform. An interpretation that would predict that pointers can have more than one position, that a cat can be both dead and alive, etc., would be a non-starter from Bohr's point of view. So the measurement problem in its usual form does not exist; it is dissolved.'' \cite[p. 24]{Dieks16}}
\end{quotation}

These two problems make explicit how QM has turned into a ``quantum omelette'' ---as Jaynes clearly expressed--- with no clear limit between the subjective and the objective, between an ontological account and an epistemological one. Both these two problems attempt to find a way out of the paradoxical mixture of an intruding subject within a supposedly objective quantum mechanical description. Let us explain this.

In the case of the basis problem, if the ``choice'' is not physically justified in terms of an objective process, the definition of reality given by the subset of properties that are actual is obviously subjective. The context is not determined {\it prior} to the choice of the experimenter and thus, it cannot be considered as {\it preexisitent}. In the measurement problem, the mix of subjective and objective pops up in the choice of the recording of an experiment ---as Wigner clearly exposed with his friend \cite{WZ}. The problem comes with the shift of the physical representation provided when the measurement was not yet performed ---and the system is described in terms of a quantum superposition (of, in the most general case, contradictory properties)---, to the single observation of a measurement outcome ---which is not described by the theory. Since there is no physical representation of the process of measurement, the choice of the recording implies a subjective aspect within the physical representation itself. Before observing the result of a measurement, it seems I can still claim that the state is in a superposition. Because there is no clear moment in which the famous ``collapse'' takes place, it becomes part of the choice of the subject to determine such a strange instant of time. Since there is no physical representation of the collapse, the subject (or his friend) seems to define it explicitly. 

However, it was not Bohr who should be considered as the main responsible for the creation of the ``quantum omelette''. Bohr never discussed questions related to the ontological nature of the quantum realm ---even though sometimes he was forced to do so.\footnote{For example, the following quotation of Bohr is also a subtle part of the omelette: ``In our description of nature the purpose is not to disclose the real essence of the phenomena but only to track down, so far as it is possible, relations between the manifold aspects of our experience.'' A ``description of nature'' is simply not the same as the ``relations between the manifold aspects of our experience.'' While the former implies ontological account, the latter assumes an epistemological one.} It was Heisenberg alone who, when introducing the so called ``Copenhagen interpretation'' in his famous book {\it Physics and Philosophy} \cite{Heis58}, mixed incoherently the epistemological complementarity scheme of Bohr with his own ontological (Platonist) approach ---which assumed a direct relation between mathematical equations and reality itself (see e.g., \cite[p. 99]{Heis71}). Heisenberg was not willing to give up on Bohr's epistemic notion of complementarity, but he was neither going to abandon the ontological problem of reality. Thus, he would support Bohr's exclusive necessity of classical physical language, and at the same time, he would argue in favor of his closed theory approach which stressed the need of creating, for each new theory, new physical concepts (see \cite{Bokulich04}). It is this deep inconsistency, in one of the major figures of the 20th Century quantum revolution, which remains at the origin of the most paradoxical present situation within the debates about quantum theory. This mixture was extended in the following decades, specially in relation to the measurement and basis problems earlier addressed.

\section{FAPP: Scrambling Ontological Problems with\\Epistemological Solutions}

One of the main constituents of the present quantum omelette is the idea that `measurement' is a process which has a special status within QM. Bohr himself made special emphasis on the idea that the analysis of the quantum measurement process was the key to recover a ``rational account of physical phenomena'' \cite{BokulichBokulich}. The `subject', `agent' or `user', must define through his choice which particular context (between the many incompatible ones) can become real (actual) ---while the others will still remain in a possible or potential realm, not truly part of physical reality, waiting for their turn to become real in the future. Bohr understood the consequences of this interpretational maneuver. And this is why he stressed repeatedly that the epistemological lesson that we must learn from quantum theory is that ``we are not only spectators but simultaneously actors in the great drama of existence''. 

The principle of decoherence introduced by Wojciech Zurek in 1981 attempted to provide an objective physical explanation of the path from the quantum formalism to ``classical reality''. Indeed, the emergence of the classical from the quantum, also known as ``the quantum to classical limit'', had remained one of the most important  open problems within the so called Copenhagen interpretation of QM. Indeed, would there be an objective explanation of such physical process that turns ``quantum particles'' into ``stable macroscopic objects'', then the subjective choices that define reality ---through the choice of the basis and the observation of the outcome--- could be finally erased from within the theory of quanta. Such objective explanation would then provide the key to a unified representation of physical reality. 

The problem of the quantum to classical limit is an ontological problem. It attempts to provide a physical explanation of {\it what is} the relation between the quantum realm and the classical realm ---both of which are presupposed to be physically real. The limit must be described in terms of a physical representation, for it seeks to explain what is going on beyond measurement outcomes and abstract mathematical formalisms.  Against the attempts of many, this problem cannot be understood in epistemological terms alone, for if there is no reference of the theory to ``something'' happening ``within physical reality'', beyond the here and now, the question regarding the path from the quantum to the classical has no definite reference, it becomes meaningless. The problem is not how a subject might acquire knowledge of the path, but rather how to {\it represent} the path in physical terms ---i.e., as physicists have always done, both mathematically and conceptually. If we assumed, from an epistemological perspective, that physical theories provide an economical account of experience with no metaphysical referent whatsoever, then there seems to be no interesting limit or relation to analyse between such theories. Both theories (within their specific limit of applicability) already accomplish their means in terms of empirical findings, and that is all there is. It simply makes no sense to talk about a limit between empirical outcomes which have no reference in the world. In conclusion, the search of the physical explanation of the path from he quantum to the classical must be regarded as a strictly representational enterprise.  

The problem of the quantum to classical limit played a major role within the Bohrian approach. It is also a problem that has helped to slowly re-cook the omelette of ontology and epistemology up to the present. Indeed, the question, when analyzed in terms of the so called ``Copenhagen interpretation'', becomes not only unclear, but simply incoherent. If QM does not relate to physical reality, if there is no conceptual representation of the quantum formalism ---as Bohr used to claim---, how could we possibly argue that there exists a limit that can be explained within physical reality?\footnote{In this respect the Bohrian solution seems to have dissolved the physical reality of one of the relata, namely, QM. This becomes clear from the fact also in the case of Bohr the formalism is regard as an abstract mathematical scheme with no direct reference to a physical representation of reality.} Clearly, the limit is not only a mathematical problem which seeks to relate two incompatible formalisms. It is also a physical problem which should be capable of providing a physical explanation of such a seemingly incompatible relation. The explanation of a physical relation between theories necessarily implies the understanding of the relata. 

For some time it was claimed by ``the new orthodoxy'' ---as Jeffrey Bub has called the followers of decoherence \cite[p. 212]{Bub97}--- that the principle of decoherence had solved the problem of the quantum to classical limit. As argued by Zurek:

\begin{quotation}
\noindent {\small``[Classical] reality emerges from the substrate of quantum physics: Open quantum systems are forced into states described by localized wave packets. They obey classical equations of motion, although with damping terms and fluctuations that have a quantum origin. What else is there to explain?'' \cite[p. 20]{Zurek02}} \end{quotation}

\noindent However, when decoherent theorists reflect about the physical meaning of such isolated quantum states, things become quite bizarre:

\begin{quotation}
\noindent {\small ``If the unknown state cannot be found out ---as is indeed the case for isolated quantum systems--- then one can make a persuasive case that such states are subjective, and that quantum state vectors are merely records of the observer's knowledge about the state of a fragment of the Universe (Fuchs and Peres 2000). However, einselection is capable of converting such malleable and `unreal' quantum states into solid elements of reality.'' [{\it Op. cit.}, p. 22]} \end{quotation}

\noindent So it seems, as some kind of powerful God, decoherence is able to create the ``real'' from the ``unreal''. 

\begin{quotation}
\noindent {\small``Quantum state vectors can be real, but only when the superposition principle ---a cornerstone of quantum behavior--- is `turned off' by einselection. Yet einselection is caused by the transfer of information about selected observables. Hence, {\it the ontological features of the state vectors ---objective existence of the einselected states--- is acquired through the epistemological `information transfer.'}'' [{\it Op. cit.}, p. 22] (emphasis added)} \end{quotation}
 
\noindent According to Zurek, decoherence is a solution to the quantum to classical limit. Ontic existence is created from epistemic choices. The subjective transfer of information of rational agents creates physical reality. If this was true, decoherence might have also solved as a corollary the very old (philosophical and religious) problem of {\it creation ex nihilo}.  

Unfortunately, irrespectively of those very strong claims, it was soon recognized that the promise to account for the quantum to classical limit was not physically justified. In fact, the principle of decoherence has been found to have many problems, {\it ad hoc} moves as well as unjustified conceptual and mathematical shortcuts. Perhaps the most important unjustified mathematical maneuver ---which we have not found analyzed in the literature--- is the ``jump'' from the (quantum) discrete to the (classical)  continuum;  i.e. the fact that an infinite numerable sum of  Hamiltonians of elementary harmonic oscillators with natural frequencies\footnote{It should be remarked that the physical meaning of an harmonic oscillator in the context of quantum theory is not at all clear. The notion of harmonic oscillator has a clear meaning in classical physics; however, the exportation of this notion to the quantum realm in not at all ``self evident''. If QM does not describe `particles' nor `waves', what is then oscillating?} is not {\it the same} as an integral of the Hamiltonians of a continuum of oscillators with real frequencies. It is this {\it ad hoc} jump from a sum to an integral which hides in itself the {\it ad hoc} imposition of classicality within QM. That which needed to be physically explained ---i.e., the path from the discrete quantum description to the continuous space-time classical description--- is simply formally imposed by Zurek \cite{Zurek81, Zurek82}. The ``continuum bath'' of harmonic oscillators is added and justified by the ``self evident'' existence of the classical realm. As it is well known, there is no decoherence when considering an infinite sum of harmonic oscillators, the path to the continuum is a necessary condition for the principle to work. However, one should be able to explain the path from the quantum discrete to the classical continuum in a clear manner just in the same way one can explain, for example, the path from classical mechanics to classical statistical mechanics. The justification of such jump is addressed by decoherent theorists arguing that all ``real systems'' are in fact, ``open systems''. The idea is that `closed systems' are ``less real'' that `open systems'. This naive understanding representation escapes the fact that `open systems' are also physical representations of reality and not reality {\it as it is}. 

In addition to these unjustifiable formal jumps and {\it ad hoc} maneuvers in the physical description, there are many other technical aspects which show the failure of the original project of decoherence to solve the quantum to classical limit. The fact that the diagonalization is not complete, since ``very small'' is obviously not ``equal to zero''.\footnote{Notice that within an epistemological account, ``very small'' might be considered as superfluous when compared to ``very big''; however, this is clearly not  the case from an ontological account. From an ontological perspective there is no essential difference between ``very big'' and ``very small'', they both have exactly the same importance.} The fact that the diagonalization can recompose itself into un-diagonalized mixtures if enough time is considered \cite{Recoherence, CormikPaz08}. The fact that the principle turns (non-diagonal) improper mixtures into (``approximately'' diagonal) improper mixtures which still cannot be interpreted in terms of ignorance.\footnote{The late recognition of this fact by Zurek has lead him to venture into many worlds interpretation in which case there are also serious inconsistencies threatening the project \cite{DawinThebault15}.} Ruth Kastner has even pointed out quite clearly why ---even if these many points would be left aside--- the main reasoning of the decoherence program is circular \cite{Kastner14}. 

There is growing consensus, mainly within the specialized literature, that the principle of decoherence fails to provide a convincing physical explanation of the quantum to classical limit.  As remarked by Guido Bacciagaluppi in his Stanford Encyclopedia of Philosophy entry on decoherence: 

\begin{quotation}
\noindent {\small``[some physicists and philosophers] still believe decoherence would provide a solution to the measurement problem of quantum mechanics. As pointed out by many authors, however (e.g. Adler 2003; Zeh 1995, pp. 14-15), this claim is not tenable. [...] Unfortunately, naive claims of the kind that decoherence gives a complete answer to the measurement problem are still somewhat part of the `folklore' of decoherence, and deservedly attract the wrath of physicists (e.g. Pearle 1997) and philosophers (e.g. Bub 1997, Chap. 8) alike.'' \cite{Bacc12}} \end{quotation}

Today, there seems to be more questions than answers when considering the solution provided by decoherence to the quantum to classical  limit (see \cite{Bacc13, Zhe96}). So it seems, the principle of decoherence might be regarded, at best, as a proto-principle, but never ---at least at this stage of its development--- as a coherent physical representation of the path from the quantum to the classical. This is not to undermine the importance of decoherence. One might recall that proto-theories have been of great importance in the development of physical theories. In QM, we find a very good example of the importance of proto-theories when recalling the Bohrian model of the atom. This model was strongly criticized, regardless of its empirical success, by both Pauli and Heisenberg due to the incoherent physical description it provided. It was the rejection of this proto-model which made necessary the requirement to develop a more general formal scheme ---later on made possible through the development of Heisenberg's matrix mechanics and Schr\"odinger's wave mechanics. In this sense, Heisenberg and Pauli's recognition of the failure of Bohr's model to produce a coherent formal and conceptual account might be regarded as the very condition of possibility for the development of QM itself. 

But after more or less having recognized that decoherence did not solve that which it had promised to solve originally, something very strange happened. Instead of reconsidering the problem and the set of presuppositions involved in order to re-develop the principle of decoherence, a new justification was advanced by this new orthodoxy. Even though it was accepted that decoherence did not ``really'' solve the quantum to classical limit, it was argued that the principle of decoherence solved the problem ``For All Practical Purposes'' (shortly known as FAPP). This was a way to claim ---more or less in disguise--- that ``we cannot really explain the path in physical terms, but don't worry, it works anyhow!'' This instrumentalist justification escapes any physical explanation and grounds itself, once again, on the predictive power of the theory ---a predictive power no physicist had ever doubted. One could argue, following an epistemic view, that decoherence is an ``economy of experience'', that it helps us to work in the lab, that it is in fact an ``epistemic solution''. But then the original ontological problem has been abandoned and replaced by an epistemic problem that makes no sense, for ---as we argued above--- there can be no interesting epistemic account of the quantum to classical (ontic) limit. 

This new ``FAPP solution'' ---which is in itself a revival of instrumentalism placed in the heart of realist discussions--- has been imported in order to discuss the ontological question of the quantum to classical limit, not only without a proper recognition of the failure of the original project but also escaping the original problem. This re-cooked omelette produced by epistemic attempts to justify decoherence ---which like many interpretations today create physical reality from epistemic choices--- has penetrated physics so deeply that today many physicist seem to uncritically accept there exists a physical process called ``decoherence'' that ``really'' takes place in the lab. However, when asked to explain what is exactly this physical process about, the new orthodoxy rapidly shifts the debate and using this new instrumentalist justification argues that: ``decoherence works FAPP!''   

The main problem of the `FAPP solution' is that it solves nothing, it just sweeps the (quantum) dirt under the (classical) carpet. By repeating that ``it works!'', many physicist and philosophers believe today that the problem has been actually solved. And there is nothing less interesting in physics than engaging in a problem that already has a solution. There is nothing there to be done, nothing to be thought or in need of development. But problems are the true gas of science, they are that which scientist work on, concentrate with passion, allow us to produce new physical theories. Problems in science should not to be regarded as ghosts or monsters that we need to destroy. There is nothing more interesting, more encouraging than a good difficult scientific problem. In QM, it was Heisenberg's and Pauli's insatisfaction with Bohr's model of the atom which led them to develop matrix mechanics and the exclusion principle. 

Instead of confronting the problem, the new orthodoxy has advanced the most weird type of justifications: ``more or less solved'', ``approximately solved'', ``almost solved'' or ``solved FAPP''. We believe it is of outmost importance to recall that {\it doxa} is not {\it episteme}, that {\it truth} and {\it falsity} cannot be equated nor regarded as ``approximate''. These are notions which possess a deep meaning, one that goes all the way back to the Greeks and the origin itself of both physics and philosophy ---as we already discussed in section 1. There is no such thing in physics as a ``more or less solved problem''. If one discusses an ontological problem ---which assumes implicitly a realist perspective regarding the physical representation of theories---, then there is no sense in talking, in epistemological terms, about ``a solution FAPP''.

\section{QBism: Unscrambling the Omelette}

The path laid down by Bohr was taken to its most extreme limit by Christopher Fuchs and Asher Peres when they wrote in the year 2000 ---exactly one Century after the beginning of the quantum voyage--- in a paper entitled {\it Quantum Theory Needs no `Interpretation'}:

\begin{quotation}
\noindent {\small ``[...] quantum theory does not describe physical reality.
What it does is provide an algorithm for computing probabilities for
the macroscopic events (`detector clicks') that are the
consequences of experimental interventions. This strict definition
of the scope of quantum theory is the only interpretation ever
needed, whether by experimenters or theorists.'' \cite[p. 70]{FuchsPeres00}}
\end{quotation}

\noindent This approach was further developed by Fuchs and R\"udiger Schack following the Bayesian interpretation of probability in order to account for QM \cite{QBism13, QBism15, QBism15b}. Because of this Bayesian perspective in order to understand QM they called their approach ``Quantum Bayesianism'', or in short:  QBism. As they remark: ``QBism agrees with Bohr that the primitive concept of experience is fundamental to an understanding of science.'' But, contrary to Bohr, ``QBism explicitly takes the `subjective' or `judgmental' or `personalist' view of probability''. In recent years David Mermin has become also part of the QBist team, publishing several papers which not only support, but also make clear the connection of QBism to the Bohrian interpretation of QM (see  \cite{Mermin04, Mermin14, Mermin14b, Mermin14c}). 

Regardless of our philosophical distance with respect to QBism, we believe that this ``no-interpretation'' is one of the most honest, consistent and clear approaches to QM. This is also the reason why QBism might allow us to begin to unscramble at least part of the quantum omelette. In this respect, maybe the most important point made by QBism is their explicit recognition of the epistemic stance they attempt to defend. The consistency of their approach is secured by their explicit denial of the existence of any (ontological) relation between QM and physical reality. QBism is a radical subjectivist approach. 

\begin{quotation}
\noindent {\small ``QBism explicitly takes the `subjective' or `judgmental' or `personalist' view of probability, which, though common among contemporary statisticians and economists, is still rare among physicists: probabilities are assigned to an event by an agent and are particular to that agent. The agent's probability assignments express her own personal degrees of belief about the event. The personal character of probability includes cases in which the agent is certain about the event: even probabilities 0 and 1 are measures of an agent's (very strongly held) belief.'' \cite[p. 750]{QBism13}}
\end{quotation}

As remarked by Fuchs, Mermin and Shack [{\it Op. cit.}, p. 750]: ``A measurement in QBism is more than a procedure in a laboratory. It is any action an agent takes to elicit a set of possible experiences. The measurement outcome is the particular experience of that agent elicited in this way. Given a measurement outcome, the quantum formalism guides the agent in updating her probabilities for subsequent measurements.'' Indeed, as QBist make explicitly clear: ``A measurement does not, as the term unfortunately suggests, reveal a pre-existing state of affairs.'' Measurements are personal, individual and QM is a ``tool'' for the ``user''  ---as Mermin prefers to call the ``agent'' \cite{Mermin14}. Just like a mobile phone or a laptop, QM is a  tool that we subjects use in order to organize our experience.

\begin{quotation}
\noindent {\small ``QBist takes quantum mechanics to be a personal mode of thought ---a very powerful tool that any agent can use to organize her own experience. That each of us can use such a tool to organize our own experience with spectacular success is an extremely important objective fact about the world we live in. But quantum mechanics itself does not deal directly with the objective world; it deals with the experiences of that objective world that belong to whatever particular agent is making use of the quantum theory.'' [{\it Op. cit.}, p. 751]}
\end{quotation}

\section{(Ontological) Problems for QBism?}

Very recently, QBism has been attacked by several authors on grounds which we will attempt to show are not tenable. Their blows not only do not touch QBism at all, but are ---at least some of them--- still grounded on the just mentioned ``quantum omelette''. Even though QBism has made clear their epistemic stance, the attacks come either from the reintroduction of ontological problems ---problems which QBism has already made very clear is simply not interested in---, or from the unwillingness to understand the radicalness of the QBist proposal. In the following we will analyze in some detail the recent criticisms to QBism presented by Mohrhoff, Marchildon and Nauenberg in a series of recent papers.

\subsection{Mohrhoff' Critics}

In \cite{Mohrhoff14} Ulrich Mohrhoff has criticized QBism for misunderstanding Bohr and the Kantian notion of objective reality. In his paper, Mohrhoff argues in favor of the soundness of ---his understanding of--- the (neo-Kantian) Bohrian project. We agree with Mohrhoff about the direct relation between Bohr and the philosopher of K\"onigsberg;\footnote{See for a detailed analysis of the relation of Bohr's philosophy to neo-Kantism \cite{Kauark}.} however, we also believe that QBism has seen much better than Bohr himself the difficult problems involved when applying an epistemological stance to understand QM. 

Mohrhoff analysis is based on the distinction of two different notions of reality, `trascendental reality' and `objective reality':\\ 

\noindent {\bf Transcendental Reality:} {\it A reality external to the subject, undisclosed in experience, which Kant looked upon as the intrinsically unknowable cause of subjective experience.} \\

\noindent Within the Kantian scheme, transcendental reality amounts to reality {\it as it is}, ``the thing in itself''. Within the Kantian architectonic trascendental reality is that which will always remain necessarily {\it veiled} ---to use a term made popular by Bernard D'Espagnat--- to the physicist.\footnote{Jacobi's famous remark makes clear the problem:  ``Without the presupposition [of the `thing in itself,'] I was unable to enter into [Kant's] system, but with it I was unable to stay within it.'' [1787: 223] As Schopenahuer would also later on make clear, the category of {\it causality} cannot be applied within Kant's system to noumenic or trascendental reality, which lies of course beyond categorical representation and (objective) experience.}\\

\noindent {\bf Objective Reality:} {\it A product of a mental synthesis based on the spatiotemporal structure of experience, achieved with the help of spatiotemporal concepts, and resulting in an objective reality from which the objectifying subject can abstract itself.}\\ 

\noindent This is, according to Kant, the reality which physicists must consider and concentrate in. A reality related to objective (represented) phenomena. While trascendental reality is an {\it absolute} notion, objective reality is a {\it relative} notion, categorically constrained and shaped by the subject. Two important points should be remarked. Firstly, the fact that the categories are subject-dependent does not imply that the subject has a saying in what objective reality amounts to. The subject can totally abstract (or detach) himself from objective reality. Secondly, the fact that the notion of objective reality which Kant puts forward within his architectonic is of course much weaker that that of {\it physis}, a notion which ---for the Greeks--- acted as the fundament of the existence and reality of Nature itself.

Following Kant, it is argued by Mohrhoff that physics should forget about `trascendental reality' and only refer to `objective reality'. However, as we shall argue, it is simply not true that QM can be considered in terms of `objective reality' within the Bohrian scheme. The subject cannot abstract himself from the definition of reality provided by QM in terms of waves, particles or even definite valued properties (see for a detailed analysis \cite{deRonde15, deRonde16}). And this is the reason why Bohr ---most of the time--- abandons any attempt to provide an account of the formalism of QM in terms of an objective representation of physical (quantum) reality. As we have already discussed, the notion of complementarity violates explicitly counterfactual reasoning which is a {\it necessary condition} for objective physical representation itself. To see this point even more clearly, imagine we have a typical double-slit set up with two subjects in front of it. The question is: can the subjects abstract themselves from the physical description? Turning the description in this way `objective'. Contrary to any experiment in classical physics or even in relativity theory, the answer ---given the Bohrian metaphysical premise according to which the description must be given in terms of classical physics by `waves' or `particles'--- is negative. Let us assume that `subject 1' chooses to measure with the two slits open and `subject 2' chooses to measure with only one slit open. If they would perform the experiment `subject 1' would conclude that the quantum object is a `wave' while ---on the contrary--- `subject 2' would conclude it is a `particle'. But an object cannot be both a `wave' and a `particle' simultaneously. At least not, if reality is considered as ``objective'', i.e. independent of particular subjective choices (see for a detailed discussion \cite{deRonde16b}). 

The real (objective) existence of waves and particles cannot be dependent on a (subjective) choice of an experimenter. In this respect, the Bohrian ``complementarity solution'' obviously precludes an objective physical representation of the quantum formalism. It is interesting in this respect to call the attention to an anecdote between Bohr and Pauli, vividly recalled by Kalervo Laurikainen. 

\begin{quotation}
\noindent{\small ``It is not generally known that there was a profound
difference in the philosophical attitudes of Niels Bohr and Wolfgang
Pauli (Laurikainen 1985b, section 3). In his address at the Second
Centenary of Columbia University in 1954, `The Unity of
Knowledge', Bohr claimed that the observer even in quantum
mechanics can be considered `detached' provided we understand the
observation in the right way (Bohr 1955, p. 83). An observation
includes a detailed description of all the experimental arrangements
which can have an influence upon the phenomenon under investigation,
and it is finished only when a registered result is obtained which
everybody can verify afterwards. In this sense, Bohr said, an
observation is quite {\it {\small objective}} (which for Bohr means
`intersubjective'), and the observer does not have any influence on
the result in any other way than by choosing the method of
observation. The result is explicitly associated with a given method
of observation. If physics is understood as a system which makes it
possible to govern such objective observational results ---which,
however, is only possible in probabilistic sense--- then physics,
according to Bohr, can even in atomic physics be considered quite
objective and the observer is `detached' in exactly the same way as
in classical physics.'' \cite[p. 42]{Laurikainen98}} \end{quotation}

\noindent Bohr had sent the manuscript of the paper to Pauli in order to
receive his critics and comments. In his reply, dated February 15,
1955, Pauli pointed out explicitly ---in line with Einstein--- that
the role of the observer in classical mechanics is essentially
different from that in quantum theory. There is no ``detached observer'' in QM. Or, in order words, the subject cannot abstract himself from the actual reality described by QM. 

\begin{quotation}
\noindent{\small ``[...] it seems to me quite appropriate to call the
conceptual description of nature in classical physics, which
Einstein so emphatically wishes to retain, `the ideal of the
detached observer'. To put it drastically the observer has according
to this ideal to disappear entirely in a discrete manner as hidden
spectator, never as actor, nature being left alone in a
predetermined course of events, independent of the way in which
phenomena are observed. `Like the moon has a definite position'
Einstein said to me last winter, `whether or not we look at the
moon, the same must also hold for the atomic objects, as there is no
sharp distinction possible between these and macroscopic objects.
Observation cannot {\it create} an element of reality like position,
there must be something contained in the complete description of
physical reality which corresponds to the {\it possibility} of
observing a position, already before the observation has been
actually made.' I hope, that I quoted Einstein correctly; it is
always difficult to quote somebody out of memory with whom one does
not agree. It is precisely this kind of postulate which I call the
ideal of the detached observer.

In quantum mechanics, on the contrary, an observation {\it hic et
nunc} changes in general the `state' of the observed system, in a
way not contained in the mathematical formulated {\it laws}, which
only apply to the automatical  time dependence of the state of a
{\it closed} system. I think here of the passage to a new phenomenon
of observation which is taken into account by the so-called
`reduction of the wave packets'. As it is allowed to consider the
instruments of observation as a kind of prolongation of the sense
organs of the observer, I consider the impredictable change of the
state by a single observation ---in spite of the objective character
of the results of every observation and notwithstanding the
statistical laws of frequencies of repeated observation under equal
conditions--- to be \emph{an abandonment of the idea of the
isolation (detachment) of the observer from the course of physical
events outside himself.}

To put it in nontechnical common language one can compare the role
of the observer in quantum theory with that of a person, who by his
freely chosen experimental arrangements and recordings brings forth
a considerable `trouble' in nature, without being able to influence
its unpredictable outcome and results which afterwards can be
objectively checked by everyone.'' 
\cite[p. 60]{Laurikainen88} (emphasis added)}
\end{quotation}

Bohr's impossibility to make sense of the quantum formalism in terms of a coherent objective description of physical reality led him very soon to consider QM only as an abstract algorithm which accounted for classical phenomena (see for discussion \cite{BokulichBokulich}). However, in many papers Bohr could not escape the need to make reference to physical reality and objectivity. That is the reason why he ended up redefining objectivity in terms of {\it intersubjectivity} ---i.e. the mutual communication of observations by rational agents.\footnote{D'Espagnat \cite{D'Espagnat06} has called such statements: {\it weakly objective} statements. I believe that calling ``objective'' a notion which does not follow the main requirement of ``objectivity'' brings a lot of confusion rather than clarification into the debate.} 

According to the author of this paper, QBism takes Bohr's stance to its only consistent conclusion when they claim explicitly that ``QM does not talk about objective physical reality''. By making explicit their limits, QBism is also helping us to begin to disentangle the quantum omelette.  Recalling Jaynes once again: ``For, if we cannot separate the subjective and objective aspects of the formalism, we cannot know what we are talking about; it is just that simple.''

\subsection{Marchildon' Critics}

Louis Marchildon \cite{Marchildon04} describes the general epistemic view ``as a way to solve the foundational problems. It does so by denying that the (in this context utterly misnamed) state vector represents the state of a microscopic system.'' He adds: ``Rather, the state vector represents knowledge about the probabilities of results of measurements performed in a given context with a macroscopic apparatus, that is, information about `the potential consequences of our experimental interventions into nature'.'' In order to discuss epistemic approaches Marchildon considers what would be ``a world for the epistemic view.'' This question, proposed by Marchildon, contradicts the very basic stance of such epistemic perspective. As QBists make clear, according to their views: {\it there is no world (nor physical reality) related to QM}. Their basic point of departure ---like it or not--- is the very denial of such an ontological {\it relation}. 

Regardless of the their standpoint, Marchildon reintroduces the ontological debate within epistemic approaches in the following way: 

\begin{quotation}
\noindent {\small ``All scientists today believe that macroscopic objects are in some sense made of atoms and molecules or, more fundamentally, of electrons, protons, neutrons, photons, etc. The epistemic view claims that state vectors do not represent states of microscopic objects, but knowledge of probabilities of experimental results. I suggest that with respect to atoms, electrons, and similar entities this can mean broadly either of three things: 1. Micro-objects do not exist [27]. 2. Micro-objects may exist but they have no states. 3. Micro-objects may exist and may have states, but attempts at narrowing down their existence or specifying their states are useless, confusing, or methodologically inappropriate.'' \cite[p. 1464]{Marchildon04}}\end{quotation}

\noindent So after accepting that the ``epistemic view claims that state vectors do not represent states of microscopic objects'', Marchildon suggests to analyze the epistemic view as related to ``Micro-objects'' anyhow. Even though we agree with Marchildon about the importance of discussing about micro-objects in QM ---because we are realists---, we regard this as an inappropriate methodological analysis, for it does not accept the premisses of the perspective under study. If ``micro-objects'' ---according to the epistemic view--- is not what QM is talking about, then it seems beyond the proposed scheme to discuss about the existence or nor of such micro-objects as related to the same theory. 

In \cite{Marchildon15}, Marchildon addresses more specifically QBism and asks the interesting question: ``How can an argument resting on personal preferences eventually move a QBist?'' And then, as in \cite{Marchildon04}, he reintroduces once again the question about micro-objects or quantum particles. He continues: ``most QBists do not deny the existence of quantum particles (i.e. electrons, photons, etc.). They deny that quantum particles have states, or that these states should be the object of science.''  Marchildon poses three different answers to the question forcing QBism to discuss about (ontic) micro-objects. This debate is clearly meaningless from a radical epistemic perspective like the one explicitly assumed by QBism. 

It is quite simple, if QBists are to remain consistent, they must remain silent about the existence of `quantum particles' simply because making reference to them would imply the idea that QM makes reference to the micro-world ---something we have seen QBists strongly deny. One could of course argue that `quantum particles' are a necessary notion within QM, but that would be a completely different problem which QBists wouldn't need to address.

\subsection{Nauenberg's Critics}

The criticisms by Michael Nauenberg discussed in \cite{Nauenberg15} have been responded by Fuchs, Mermin and Schack in \cite{QBism15}. However, we would like to stress the fact that the conclusion of Nauenberg makes clear he is not discussing QBism at all. Nauenberg argues that: ``Contrary to Fuchs et al., quantum theory deals with the objective world as directly as does classical mechanics.'' A statement of such certainty makes clear that realism should be not understood as a ``belief'', but as a perspective. This is not an argument, it is rather stating a philosophical position and claiming that a different one is simply false. 

In the last sentence of their reply to Nauenberg, QBists implore for a true debate within QM. I agree with their remark: ``We welcome criticism, but urge critics to pay some attention to what we are saying.'' Someone saying ``I am a realist, and you are wrong for being a QBist'' is not producing any debate, that is simply begging the question.

\subsection{In Defense of QBism}

Morhof, Marchildon and Nauenberg do not respect the premises of QBism, once and again they ask QBists to answer ontological questions they have explicitly left aside right from the start. QBists ---like Bohr--- are not interested in discussing about physical reality. In fact, as we have seen above, QBism stands on the very denial of the relation between objective reality and QM. 

Philosophical perspectives or positions limit explicitly the questions that can be made, the problems that can be addressed. Of course, one might disagree with a particular stance or viewpoint and argue against it. But we should be aware that what produces no results whatsoever is to argue without respecting the presuppositions of our opponent. One cannot ask an instrumentalist to be a realist, or an empiricist to become an idealist, or a QBist to believe that quantum theory talks about the world and reality. A philosophical stance is a guiding line, not something that can be tested in the lab. What really matters in philosophical and foundational debates is the consistent and critical reasoning of ideas and presuppositions, something we have almost lost in the discussions about QM. Exactly this is something which Qbism is helping us to recover.

\section{QBism: ``Solving'' or ``Dissolving'' (Ontological) Problems?}

QBism makes clear its own stance, it makes explicit the presuppositions it stands for. This honesty can help us to unscramble at least part of the quantum omelette. By denying any ontological or objective reference of QM to the world and physical reality QBism rejects ontological debates right from the start. This move allows QBism to escape all ontic-problems of QM: the measurement problem, the basis problem, non-locality, non-separability, non-identity, etc. They need not get into these difficult problems because according to them QM simply does not make reference to objective reality. QM works FAPP! And that is enough for QBists. 

Within the limits imposed by their radical epistemic perspective, all this is perfectly consistent. However, the strength of QBism, namely, its denial of the relation of QM to physical reality, is also its weakness. Because of this denial, QBism is not falsifiable, it simply cannot be proven to be wrong. Because it says nothing at all about objective reality, the only way to evaluate QBism is in relation to the personal observation of subjects which bet in a quantum casino using a quantum algorithm. But because subjective observations are personal, by definition, there is no possibility of comparing observations in objective ---subject independent--- terms. 

QBism is an economy of personal experience, but a personal experience has no reference but itself. It is restricted to each agent. QBism is consistent when it restricts to the beliefs of users gambling in a quantum casino. But gamblings and beliefs cannot be falsified. QBism is completely safe from criticisms simply because it says nothing beyond personal observation, it detaches QM from the world and reality. In this way QBists refuse to explain {\it why} `quantum clicks' appear in such a weird non-classical manner. By doing this they dissolve all important and interesting questions that physical thought has produced since the origin itself of the theory of quanta. Taking to its most extreme limit several of the main Bohrian ideas, QBism has turned physics into a solipsistic realm of personal experience in which no falsification can be produced; and even more worrying, where there are no physical problems or debates left. QBism does not solve the problems of QM, it simply dissolves them.

\section{The Importance of Ontological Problems in\\ Quantum Physics}

For QBists, we have already reached the end of the road. The problems have been (dis)solved. We should stop going to foundational conferences to discuss about QM and its relation to  the world and physical reality. For the QBist, the ``solution'' to all problems is quite simple: ``just stop asking those weird metaphysical and ontological questions!'' Indeed, QBists are right: the moment you stop, the problems immediately disappear, they vanish. However, we should not forget that it was these same problems, which were discussed all around the world since the very origin of QM ---going back to the Solvay meetings---, the same ones that made possible the development of a new amazing experimental and technological era. According to our perspective, it is in fact these ontological problems which have been the true gas of quantum mechanical developments. 

The EPR discussions about the meaning of physical reality. Bell's inequalities, which discuss EPR type experiments and the possibility to represent them through a local-realistic model or hidden variable theory. The 1935 {\it gedankenexperiemnt} of Schr\"odinger which exposes the impossibility to imagine quantum superpositions and entanglement in classical terms ---something Paul Dirac had already stressed in 1930 within his famous book {\it Principles of Quantum Mechanics}. The Kochen-Specker theorem, which analyzes the limits to the interpretation of projection operators in terms of actual (definite valued) preexistent properties. All these problems and questions presuppose as a standpoint a realist perspective according to which {\it QM does relate to physical reality}. It is these problems and questions the ones that  allowed us to produce outstanding developments such as quantum teleportation, quantum cryptography and quantum computation. 

Today we know that Einstein and Bell were wrong to assume that a local-realistic hidden variable model would be able to reproduce the predictions of QM. We have learned about superpositions and entanglement, and we even suspect that Schr\"odinger's cat might be even fat!\footnote{See in this respect \cite{Blatter00}. In fact, it becomes increasingly clear that quantum superpositions are telling us something about quantum physical reality even at the macroscopic scale \cite{Nature15, NimmrichterHornberger13}.} Today we also know that quantum contextuality is intrinsic to the formalism of the theory and we have learned how to deal with mutually incompatible quantum contexts. All we have learnt of quantum information processing cannot be disconnected from EPR and Bell's analysis, from Schr\"odinger's reflections regarding the superpositions or  Kochen-Specker type theorems. The teachings that have been produced in the field of foundations and philosophy of QM should not be underestimated. 

We believe that physical problems need realism to grow, to flourish. Problems will never grow in the land of QBism. Instrumentalism creates a desert where no physical questions are allowed to see the light. All interesting problems which we have been discussing in the philosophy of science and foundations community for more than a Century ---problems which Fuchs thinks are a burden for the taxpayer\footnote{As argued by Fuchs in \cite{Fuchs02}: ``The issue remains, when will we ever stop burdening the taxpayer with conferences devoted to the quantum foundations?'' According to Fuchs what you find in conferences is the following: ``Go to any meeting, and it is like being in a holy city in great tumult. You will find all the religions with all their priests pitted in holy war [Bohmians, CH, MW, Everettians, etc.]  They all declare to see the light, the ultimate light. Each tells us that if we will accept their solution as our savior, then we too will see the light. But there has to be something wrong with this! If any of these priests had truly shown the light, there simply would not be the year-after-year conference. The verdict seems clear enough: If we ---i.e., the set of people who might be reading this paper--- really care about quantum foundations, then it behooves us as a community to ask why these meetings are happening and find a way to put a stop to them.''}--- have been in fact the conditions of possibility for the development of a new quantum technological era.  

QM places us in a crossroad from which we must decide which path to follow (and that makes all the difference). The foundational discussions that have taken place during the last decades are in strict relation to a realist account of the theory. So either you accept that QM talks about the world and reality, and then you have an enormous problem to face, maybe the most important problem in physics of all times: you must clearly explain what is QM telling us about the world and physical reality. Or you can also accept, following QBism, that QM does not make reference to anything but `beliefs of users' and `measurement outcomes', that the theory of quanta is in fact, only an algorithm that computes `clicks' in detectors. The choice of assuming the first path leaves you with an amazing problem to confront: we possess an empirically adequate theory which has produced the most outstanding technical revolution in the 20th Century, which I believe will change drastically our world in the 21st Century, but we still do not know what the theory is talking about in terms of physical reality. The choice of the latter, on the contrary, leaves you with no single problem to address: everything has been already solved FAPP. 

What is important to remark at this point is that you cannot transit through both ontic and epistemic paths at the same time ---as Mermin would like to do \cite{Mermin14b, Mermin14c}. You simply cannot claim simultaneously, on the one hand, that ``QM does not describe physical reality'' and on the other hand, that ``there is a world which relates to QM''. These are two contradictory statements. Either there exists a {\it relation}, and then we are confronted with the problem of trying to find out what {\it is} the nature of such relationship. We need to explain the relation. Or, we could also argue, following QBism, that such a relation does not exist. There is no reference whatsoever of the theory to the world and reality. And that is simply the end of the road. 

The choice of the path is personal and deals explicitly with the assumption of a philosophical stance. I myself like problems, I find them encouraging, I see them as guiding lines that allow us to keep going. However, whatever choice is made we should be careful not to add more confusion to the quantum omelette already created and still cooked in many papers today. Such omelette, as we have argued above, has been developed from the exclusive discussion and analysis of ---mainly--- {\it pseudoproblems}.

\section{Final Remarks: Physics Back to {\it Physis}}

Since the Greeks, we physicists and philosophers, have been marveled with the possibility of understanding reality and the world. That is the origin of both physics and philosophy. We might be called dreamers, naive people, but these strange ideas have got us quite far... Sophists, on the contrary, might still believe that subjects, agents or users are the true measure of all `clicks'. Both perspectives have fought through centuries, they have become through these battles the kernel of Western thought, maybe two sides of the same coin. But we should not confuse them, for their presuppositions and questions differ in methodology and nature. Their specific problems are in themselves conditions of possibility which imply limits to reasoning. One simply cannot jump from an ontological question into an epistemological answer, as it has been done in the context of the analysis of decoherence; or from an epistemological question into an ontological answer, as we have shown do the main attacks against QBism. I believe that both epistemology and ontology are of deep importance for the development of knowledge. But we have to stop mixing the ontological and epistemological levels of analysis.

We believe that a balance between honesty to state one's own philosophical presuppositions, and bravery to stand and argue in favor of them, are a basic necessary condition for the production of meaningful academic research. In this respect, standing at the very opposite corner of QBism, we have argued in favor of grounding physics on its own original fundament: {\it physis}. QM confronts us with one of the most important problems in the history of science, we might choose to confront it, or we might also want to escape it. We are convinced that the realist has no choice but to stand the fight.  

\begin{quotation}
\noindent {\small ``The important thing is not to stop questioning. Curiosity has its own reason for existence. One cannot help but be in awe when he contemplates the mysteries of eternity, of life, of the marvelous structure of reality. It is enough if one tries merely to comprehend a little of this mystery each day. Never lose a holy curiosity. ... Don't stop to marvel.'' \cite[p. 64]{Einstein55}}
\end{quotation}

\bigskip

\section*{Acknowledgements} 

This work was partially supported by the following grants: FWO project G.0405.08 and FWO-research community W0.030.06. CONICET RES. 3646/14 (2013-2014) and the Project PIO-CONICET-UNAJ (15520150100008CO) ``Quantum Superpositions in Quantum Information Processing''.


\end{document}